\documentclass[pre,aps,twocolumn,superscriptaddress,amsmath,showpacs,floatfix]{revtex4}
\usepackage{graphicx}
\begin{document}
\title{Biased transport of elastic cytoskeletal filaments with alternating polarities by molecular motors}
\author{Barak Gur}
\affiliation{Department of Physics, Ben Gurion University,
Be'er Sheva 84105, Israel}
\author{Oded Farago}
\affiliation{Department of Biomedical Engineering, Ben Gurion University,
Be'er Sheva 84105, Israel}

\begin{abstract}

We present a simple model for the bidirectional dynamics of actin
bundles with alternating polarities in gliding assays with
non-processive myosin motors. In the model, the bundle is represented
as an elastic chain consisting of monomers with positive and negative
polarities. The motion of the bundle is induced by the pulling forces
of the underlying motors which stochastically attach to the monomers
and, depending on the polarity of the monomers, pull them in the right
or left direction. We demonstrate that perfectly a-polar chains
consisting of equal numbers of monomers with positive and negative
polarities may exhibit biased bidirectional motion with non-zero
drift. This effect is attributed to the elastic tension developed in
the chain due to the action of the myosin motors. We also show that as
a result of this tension, the attachment probability of the motors is
greatly reduced and becomes strongly dependent on the length of the
chain. These surprising effects point to the necessity of considering
the elasticity of the cytoskeleton in theoretical studies of
cooperative dynamics of molecular motors.

\end{abstract}
\maketitle

Motor proteins are molecular machines that convert chemical energy
into mechanical work by ATP hydrolysis. They ``walk'' on the
microtubule and actin cytoskeleton and pull vesicles or organelles
across the cell \cite{alberts}. The intracellular transport of cargoes
is achieved mainly by the action of individual motors which propagate
along the cytoskeleton tracks in a direction determined by the
intrinsic polarity of the filaments \cite{kreis}. Other processes,
such as cell motility and mitosis, require the cooperative work of
many motors. Muscle contraction, for instance, involves the
simultaneous action of hundreds of myosin II motors pulling on
attached actin filaments and causing them to slide against each other
\cite{geeves}. One interesting outcome of cooperative action of
molecular motors is their ability to generate bidirectional motion
\cite{badoual}. Bidirectional movement results from the competition
between two populations of motors that work against each other in
opposite directions \cite{badoual,klumpp,muhuri,hexner,zhang}. The
direction of motion flips from one direction to the other due to
stochastic events of binding and unbinding of motors to the filament
which tip the force balance between the two motor groups.

The dynamics of motor-filament systems are often studied using in
vitro motility assays in which the filaments glide over a dense bed of
immobilized motors and their motion is tracked by fluorescent
microscopy \cite{pollard}. Recently, we used such a motility assay to
study the dynamics of actin bundles induced by the cooperative action
of myosin II motors \cite{gilboa}. The bundles in these experiments
were composed of short actin segments which, through a sequence of
fusion events, assemble into filaments with randomly alternating
polarities. Such a-polar bundles exhibit ``back and forth''
bidirectional motion. We showed that the distribution of ``reversal
times'' (i.e., the durations of unidirectional intervals of motion)
take an exponential form $P(\Delta t)\sim \exp{ ({-\Delta
t}/{\tau_{\rm rev}})}$, where $\tau_{\rm rev}$ is the characteristic
reversal time of the bidirectional motion
\cite{gilboa,gillo}. Detailed analysis of the dynamics of many bundles
revealed that $\tau_{\rm rev}$ is of the order of a few second and has
no dependence on the length (number of monomers, $N$) of the
bundle. This result was in marked contradiction with previous
theoretical models which predicted that $\tau_{\rm rev}$ grows
exponentially with $N$ \cite{badoual}. To resolve this disagreement,
we have introduced a model that takes into account the elastic energy
stored in the actin bundle due to the action of the working motors
\cite{gilboa,gillo}. The elastic energy modifies the rates at which
motors attach to and detach from the actin and eliminates the
exponential dependence of $\tau_{\rm rev}$ on $N$.

Our previous theoretical treatment of cooperative bidirectional motion
was based on a mean field calculation of the actin elastic energy,
ignoring both: (i) the sequential order of the polarities of the
monomers, and (ii) the positions along the filament where the pulling
forces of motors are applied. The mean field elastic energy scales as
$E\sim NN_c$, where $N_c$ is the number of attached motors
\cite{gilboa,gillo}. Within the mean field picture, the bidirectional
motion on perfectly a-polar tracks consisting of an equal number of
monomers with right-pointing (``positive'') and left-pointing
(``negative'') polarities has no bias, i.e., the intervals of motion
in both directions occur with equal probability. In this letter we
discuss an interesting effect related to the elasticity of the
actin. We show that a-polar elastic filaments may exhibit a biased
bidirectional motion and achieve a net migration along the
motors-coated surface.  For myosin II-actin systems, we find that the
drift velocity is typically 2-3 orders of magnitude smaller than the
velocity of a single myosin II motor and is comparable with the speed
by which the motors move the a-polar bundle cooperatively during the
intervals of unidirectional motion. This newly identified mechanism of
propagation may, therefore, be relevant to processes of active
self-organization of cytoskeletal structures during which filaments
are transported and joined with each other by motor proteins.

\begin{figure}[t]
\begin{center}
\scalebox{0.5}{\centering \includegraphics{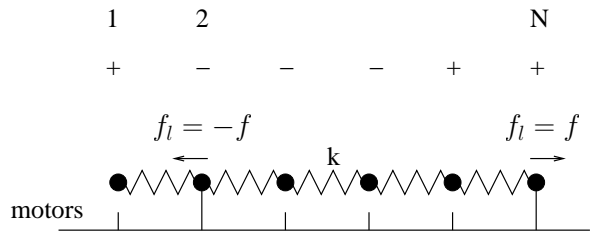}}
\end{center}
\vspace{-0.5cm}
\caption{A schematic drawing of the system: A chain of consisting of
$N$ monomers with alternating polarities, connected to each other by
identical springs. The chain lies on a ``bed'' of motors, some of
which are connected to the monomers. A connected monomer with positive
(negative) polarity feels a pulling force of size $+f$
($-f$). Disconnected monomers experience no force.}
\label{Figure1} 
\end{figure}

To demonstrate the effect, we consider the chain illustrated in
Fig.~\ref{Figure1}, consisting of $N$ monomers connected by $(N-1)$
identical springs with a spring constant $k$. Each monomer may be
either free and experience no pulling force ($f=0$), or attached to
one motor in which case it is subjected to a force of magnitude $f$
which is directed to the right ($+f$) for monomers with positive
polarities and left ($-f$) for monomers with negative polarities. The
moving velocity of the filament is given by $V= f_{\rm total}/
\lambda$, where $f_{\rm total}=\sum _{l=1}^{N}f_l$ is the sum of motor
forces applied on the monomers and $\lambda$ is the friction
coefficient of the chain. A chain of $N$ monomers has $2^N$ connection
configurations, where each such configuration can be represented by a
vector $\vec{C}$ of size $N$ specifying the state
(connected/disconnected) of each monomer. For example, a chain of 4
monomers in which the first and third monomers are connected to motors
will be represented by $\vec{C}=(1,0,1,0)$. Let us also introduce a
vector $\vec{S}$ whose components are related to the polarities of the
monomers. The vector $\vec{S}=(1,1,-1,1)$, for instance, corresponds
to a chain of 4 monomers in which the polarities of the first, second,
and fourth monomers is positive while the third monomer has a negative
polarity. The drift velocity can be calculated by averaging over all
possible connection configurations of the motors (all possible values
of the vector $\vec{C}$):
\begin{equation}
V_{\rm drift}({\vec{S}})\equiv \langle V\rangle = \sum_{j=1}^{2^N}
\frac{f}{\lambda_j}\left(\vec{C}_j\cdot\vec{S}\right)P_j,
\label{eq:drift}
\end{equation}
where $P_j$ is the occurrence probability of the configuration, and the
subscript $j$ has been added to $\lambda$ to account for possible
variations in the friction coefficient between the different
configurations. The probability $P_j$ depends on (i) the number of
attached motors in the configuration, $N_c(j)=\|\vec{C}_j\|^2$, (ii)
the attachment probability of a single motor, $q$, and (iii) the total
elastic energy of the springs $E_j^{\rm el}$:
\begin{equation}
P_j=\frac{1}{Z}q^{N_c(j)} (1-q)^{(N-N_c(j))}e^{-\beta E_j ^{\rm el}},
\label{eq:probability}
\end{equation}
where $\beta=(k_BT)^{-1}$ is the inverse temperature and $Z$ is the
partition function of the system. The elastic energy is the sum of the
energies of the springs, $E_j^{\rm{el}}=\sum _{i=1}^{N-1} F_i^2/2k$,
where $F_i$ is the force stretching (or compressing) the $i$-th
spring.  The forces $F_i$ can be calculated using the following steps:
(i) calculate the mean force $\bar{f}\equiv f_{\rm
total}/N=f(\vec{C}\cdot\vec{S})/N$, (ii) calculate the access forces
acting on the monomers $f^*_l=fC_lS_l-\bar{f}$, and (iii) sum the
access forces applied on all the monomers located on one side of the
spring $F_i=\sum_{l=1}^i f^*_l=-\sum_{l=i+1}^N f^*_l$. Our analysis is
based on the assumption that variations in $\vec{C}$ (which occur when
motors attach to or detach from the actin track) lead to instantaneous
changes in the velocity of the filament which should always be
proportional to the total exerted force. This assumption is expected
to hold for low Reynolds numbers where inertia can be neglected.

Let us analyze the dynamics of a chain of size $N=4$. There are six
different a-polar sequences for a chain of this length:
$\vec{S}_{1a}=-\vec{S}_{1b}=(1,1,-1,-1)$,
$\vec{S}_{2a}=-\vec{S}_{2b}=(1,-1,1,-1)$,
$\vec{S}_{3a}=-\vec{S}_{3b}=(1,-1,-1,1)$. It is easy to prove that the
drift velocity (see Eqs.~(\ref{eq:drift}) and (\ref{eq:probability}))
vanishes for the first four sequences which are antisymmetric with
respect to reflection around the midpoint. This is not the case with
the last two symmetric sequences. To see this, consider the sequence
$\vec{S}_{3a}$ and assume, for simplicity, that
$\lambda_j=\lambda$. In the limit $q\ll 1$, one can ignore the
configurations in which more than one out of the four monomers is
connected to a motor. This leaves us with only five
configurations: (i) $\vec{C}_j=(0,0,0,0)$, for which $V_j=0$ and
$P_j=(1-q)^4/Z$, (ii) $\vec{C}_j=(1,0,0,0)$ and $\vec{C}_j=(0,0,0,1)$,
for which $V_j=f/\lambda$ and $P_j=q(1-q)^3e^{-(7/8)(\beta
f^2/2k)}/Z$, and (iii) $\vec{C}_j=(0,1,0,0)$ and
$\vec{C}_j=(0,0,1,0)$, for which $V_j=-f/\lambda$ and
$P_j=q(1-q)^3e^{-(3/8)(\beta f^2/2k)}/Z$. Substituting this in
Eq.~(\ref{eq:probability}) gives $V_{\rm drift}(\vec{S}_{3a})=-V_{\rm
drift}(\vec{S}_{3b})\simeq -2(f/\lambda)q [e^{-(3/8)(\beta
f^2/2k)}-e^{-(7/8)(\beta f^2/2k)}]$. For $\beta f^2/2k\ll 1$ we find
that the drift velocity increases with a third power of the motor
force, $V_{\rm drift}\simeq -(\beta/2k\lambda)f^3$. This power law has
a different exponent than 1~--~the scaling exponent for the velocity
of stiff polar chains.

\begin{figure}[t]
\begin{center}
\scalebox{0.5}{\centering \includegraphics{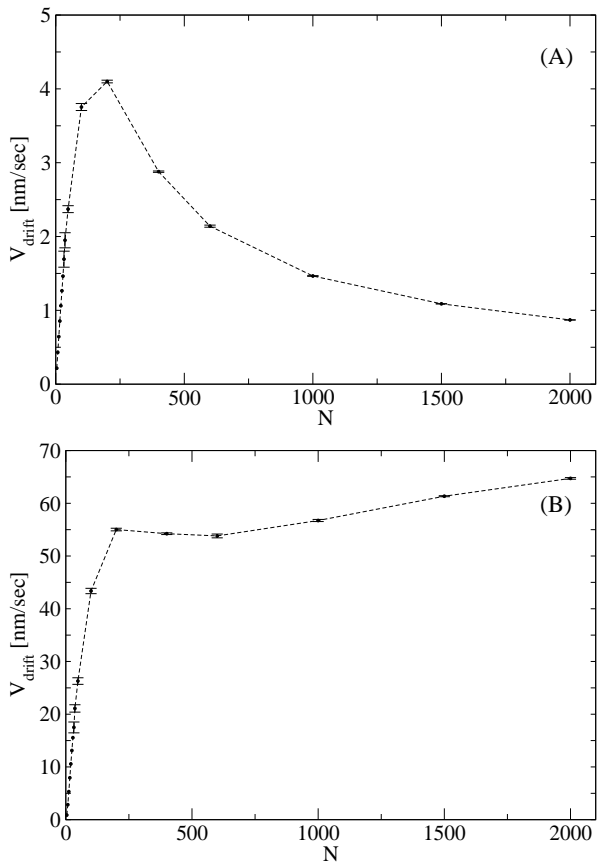}}
\end{center}
\vspace{-0.5cm}
\caption{The drift velocity $V_{\rm drift}$ as a function of the
  length of the chain. The friction coefficient $\lambda_j$ is
  proportional to the number of monomers $N$ in (A) and the number of
  connected motors $N_c(j)$ in (B). The lines are guides to the eye.}  
\label{Figure2} 
\end{figure}

\begin{figure}[t]
\begin{center}
\vspace{0.7cm}
\scalebox{0.35}{\centering \includegraphics{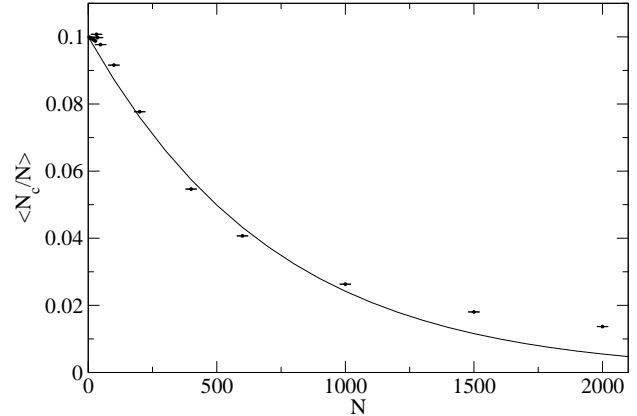}}
\end{center}
\vspace{-0.5cm}
\caption{The mean fraction of monomers connected motors, $\langle
N_c/N\rangle$ as a function of $N$. The solid line represents the mean
field result {\protect Eq.(\ref{eq:processivity}) with $c=0.75$}.}
\label{Figure3} 
\end{figure}

To further investigate this effect, we calculated the drift velocity
for chains of $N=4M$ monomers with sequences of the form
$\vec{S}=(\overbrace{-1,\ldots,-1}^M,\overbrace{1,\ldots,1}^{2M},\overbrace{-1,\ldots,-1}^M)$. Our
results are summarized in Figs.~\ref{Figure2}(A) and
(B). Fig.~\ref{Figure2}(A) is based on a calculation in which the
friction coefficient (see Eq.~(\ref{eq:drift})), $\lambda_j=\lambda_0
N$,
while in Fig.~\ref{Figure2}(B), we assumed that $\lambda_j=\lambda_0
N_c(j)$. The results for $N \leq 28$ have been derived using a full
statistical calculation of the partition function, while for larger
$N$ they have been obtained from Monte Carlo simulations. The model
parameters were assigned the following values which are representative
of myosin II-actin systems \cite{gillo,howard}: $\beta f^2/2k=0.002$,
$q=0.1$, and $f/\lambda_0=6\ {\rm \mu m/sec}$. Both figures show that
for small chains of size $N<200$, the drift velocity increases rapidly
with $N$. For larger chains ($N>200$), $V_{\rm drift}$ behaves
differently in Figs.~\ref{Figure2}(A) and (B). In the former it
decreases with $N$, while in the latter it saturates and increases
again for $N>600$. Note also that the different scales of the y-axis
in both figures. These differences can be attributed to the different
values of $\lambda_j$ used in the cases represented by
Figs.~\ref{Figure2}(A) and (B). Since for each configuration, the
ratio between the friction coefficients in both cases
$r_{\lambda}\equiv\lambda_j^B/\lambda_j^A=N_c(j)/N\leq 1$, the drift
velocity in (B) must always be larger than in (A). Fig.~\ref{Figure3}
depicts the mean value of $r_{\lambda}$ (i.e., the mean fraction of
connected monomers) as a function of $N$. For $N<50$, $\langle
r_{\lambda}\rangle\simeq q=0.1$ and, accordingly, the ratio between
the drift velocities in (B) and (A) in this regime is close to one
order of magnitude. For $N>50$, $\langle r_{\lambda}\rangle $ drops to
values much smaller than $q$, which implies that the friction
coefficient {\em per monomer}\/ decreases with $N$ in case (B) and
explains why the drift velocity remains high and does not decrease
sharply as in (A). The decrease in the mean fraction of connected
monomers can be traced to the fact that configurations with larger
$N_c(j)$ have, in general, higher elastic energies and, therefore,
smaller statistical weights. For the mean field elastic energy, $E_j
^{\rm el}/k_BT=c(\beta f^2/2k)NN_c(j)$ \cite{gilboa,gillo}, one gets
\begin{equation}
\langle r_{\lambda}\rangle=\left\langle \frac{N_c}{N}\right
\rangle=\frac{q\exp(-cN\beta
f^2/2k)}{1-q[1-\exp(-cN\beta
f^2/2k)]},
\label{eq:processivity}
\end{equation}
where $c$ is a dimensionless constant of the order of 1.  For $c=0.75$
and $N\leq 1000$, this expression (solid line in Fig.~\ref{Figure3})
gives a fair agreement with the computational results. For larger
values of $N$ (i.e., when $\langle N_c/N \rangle$ becomes very small),
the expression tends to overestimate the rate of decrease in the mean
fraction of connected monomers (or, equivalently, the effective
attachment probability).  The decrease in the attachment probability
of the motors is another, indirect, manifestation of cooperativity
between the motors which is mediated through the forces that they
jointly exert on the actin track. Eq.~(\ref{eq:processivity}) suggests
that the elasticity of the track can be neglected for small filaments
whose size $N\ll(\beta f^2/2k)^{-1}\equiv N^*$. In this regime, the
two cooperativity effects discussed here which are associated with the
elasticity of the actin filaments disappear: (i) The drift velocity
$V_{\rm drift}\sim (\beta/2k\lambda)f^3=(f/\lambda)(N^*/N)\ll
(f/\lambda)$ is vanishingly smaller than the typical speed by which
the bidirectionally moving bundle propagates in each direction, and
(ii) the fraction of attached motors $\langle N_c/N \rangle\simeq q$
is very close to the attachment probability of individual motors. The
elasticity effects can be detected only for long filaments with
$N\gtrsim N^*$, which are softer (the effective force constant of the
filament decrease as $N^{-1}$) and, hence, more influenced by the
forces of the motors. For infinitely stiff filaments ($k\rightarrow
\infty$), the crossover filament size diverges
($N^*\rightarrow\infty$) and the filament elasticity is, of course,
irrelevant on all length scales.

Figs.~\ref{Figure2}(A) and (B) represent two limiting cases. In the
former, the friction is caused by the drag of the actin bundle in the
viscous environment, while in the latter it originates from the
attachment of the actin to the underlying surface of motors. The
actual friction coefficient is expected to lie between these two
extreme values and, therefore, the drift velocity should exhibit an
intermediate behavior between those shown in Figs.~\ref{Figure2}(A)
and (B). Thus, the typical magnitude of $V_{\rm drift}$ is expected to
be of the order of 10 nm/sec. Interestingly, the drift velocity of the
bundle is of the same order of magnitude as its speed during the
bidirectional motion \cite{gilboa}, which has also been found to be
2-3 order of magnitude smaller than the moving velocity of individual
myosin II motors ($v\sim 6\ {\rm \mu m/sec}$ \cite{howard}). Over a
period of a few minutes the a-polar bundle may progress a distance of
a few micrometers. This implies that the drift of a-polar bundles may
be relevant to the active remodeling of the cell cytoskeleton
occurring during many cellular processes. 

Our investigation of the role of the filament elasticity in modifying
collective motion of molecular motors has been motivated by
experiments which have been described and analyzed by using a ratchet
model and a mean field approximation for the elastic energy
\cite{gilboa,gillo}. In this paper we presented a more realistic
microscopic based model that involves the determination of the exact
elastic energy of the filaments. We demonstrated that such a model
leads to new insights and novel results like the biased transport of
filaments with no net polarities. Experimental verification of this
surprising result is, however, difficult. It requires that (i) the
moving filaments are perfectly a-polar with internal (sequential)
order, and (ii) that they move for sufficiently long period of time
such that the net drift can be extracted from the statistics of the
unidirectional intervals of motion. Unfortunately, the a-polar bundles
are not formed by a well controlled process, but rather through a
sequence of stochastic fusion events that usually generate filaments
with disordered, random, sequences and with little residual polarities
\cite{gilboa}. Also, in the existing experimental setup, the
bidirectional motion cannot be tracked for more than about 10 minutes,
which is too short for a meaningful statistical analysis. What should
be more experimentally testable is the other elasticity effect, namely
the reduction in the fraction of connected motors. This effect, which
has been attributed to the dependence of the elastic energy on the
configuration of connected motors (denoted by the vector $\vec{C}$),
is not limited to a-polar filaments. Polar filaments experiencing a
non-uniform distribution of motor forces (i.e., when only a fraction
of the monomers are connected to motors) will also develop a tensile
stress that could potentially alter the attachment probability of the
motors. In a future publication we plan to present a theoretical
analysis of the attachment probability for perfectly polar filaments,
similar to the analysis presented here for a-polar filaments. We also
plan to investigate this effect experimentally by using a motility
assay combined with micro-manipulation technique (such as optical
tweezers) to stall the gliding filament and measure the mean force
generated by the motors. In the case of perfectly polar filaments, the
forces of all the motors are applied along the same direction and,
therefore, the total measured force should be simply proportional to
the number of attached motors.

We thank Anne Bernheim-Groswasser for numerous valuable
discussions. BG wishes to thank Amir Erez for his assistance with the
numerical work and for many interesting discussions.


\end{document}